\documentclass[prl,aps,groupedaddress,twocolumn,floatfix,nofootinbib]{revtex4}
\usepackage{amsmath,amsthm,amsfonts,amssymb,graphics,graphicx,epsfig,color,times,natbib,subfigure}
\usepackage{bbm} 

\usepackage{soul} 

\usepackage{enumitem}

\newcommand{\ket}[1]{\mbox{$ | #1 \rangle $}}
\newcommand{\bra}[1]{\mbox{$ \langle #1 | $}}

\newcommand{\be}{\begin{equation}}
\newcommand{\ee}{\end{equation}}
\newcommand{\ba}{\begin{eqnarray}}
\newcommand{\ea}{\end{eqnarray}}

\newcommand{\one}{\leavevmode\hbox{\small1\normalsize\kern-.33em1}}

\begin{document}

\title{Not all entangled states violate Leggett's crypto-nonlocality}

\author{Cyril Branciard}
\affiliation{Centre for Engineered Quantum Systems and School of Mathematics and Physics, The University of Queensland, St Lucia, QLD 4072, Australia}

\date{\today}

\begin{abstract}

This note is a reply to M. Navascu\'es' claim that ``all entangled states violate Leggett's crypto-nonlocality''~[arXiv:1303.5124v2]. I argue that such a conclusion can only be reached if one introduces additional assumptions that further restrict Leggett's notion of ``crypto-nonlocality''. If \emph{a contrario} one sticks only to Leggett's original axioms, there exist entangled states whose correlations are always compatible with Leggett's crypto-nonlocality---which is thus a genuinely different concept from quantum separability. I clarify in this note the relation between these two notions, together also with Bell's assumption of local causality.

\end{abstract}

\maketitle

\paragraph{Leggett's crypto-nonlocality.---}

The concept of ``crypto-nonlocality''~\cite{leggett} was introduced by A.~J.~Leggett in an attempt to explain quantum correlations with some kind of ``realistic'' picture: roughly speaking, it says that all individual subsystems of a composite system should locally behave as if they were in a pure quantum state, with well-defined properties.

To make it more precise, consider (following Leggett) the simplest case of bipartite correlations obtained from projective measurements on a 2-qubit state. The measurement settings can be described by unit vectors $\vec a$ and $\vec b$ on the Bloch sphere ${\cal S}^2$, and the measurement outcomes are binary variables, denoted here by $\alpha, \beta = \pm 1$. The correlation observed by the two parties, Alice and Bob, is then described by the joint conditional probability distribution $P(\alpha, \beta | \vec a, \vec b)$.
According to Leggett's assumptions, it should be possible to decompose this correlation as a mixture of correlations $P_{\vec u, \vec v}(\alpha, \beta | \vec a, \vec b)$ whose local marginal probability distributions for Alice and Bob are those corresponding to pure qubits in the states $\ket{\vec u}$ and $\ket{\vec v}$, respectively, represented by unit vectors $\vec u$ and $\vec v$ on the Bloch sphere. I.e., the correlation $P(\alpha, \beta | \vec a, \vec b)$ is compatible with Leggett's crypto-nonlocality \emph{if and only if} there exists a non-negative, normalised probability distribution $\rho(\vec u, \vec v)$ and correlations $P_{\vec u, \vec v}(\alpha, \beta | \vec a, \vec b)$ ($\geq 0$) such that
\ba
P(\alpha,\beta | \vec a, \vec b) &=& \int_{\!{\cal S}^2} \int_{\!{\cal S}^2} \! P_{\vec u, \vec v}(\alpha, \beta | \vec a, \vec b) \, \rho(\vec u, \vec v) \, d\vec u \, d\vec v , \quad \label{decomp_leggett}
\ea
with the marginals of $P_{\vec u, \vec v}(\alpha, \beta | \vec a, \vec b)$ satisfying
\ba
M^A_{\vec u, \vec v}(\vec a, \vec b) &:=& \sum_{\alpha,\beta} \alpha \, P_{\vec u, \vec v}(\alpha, \beta | \vec a, \vec b) \, = \, \vec u \cdot \vec a \label{constr_ua}, \quad \\
M^B_{\vec u, \vec v}(\vec a, \vec b) &:=& \sum_{\alpha,\beta} \beta \, P_{\vec u, \vec v}(\alpha, \beta | \vec a, \vec b) \, = \, \vec v \cdot \vec b \label{constr_vb},
\ea
for all $\vec u, \vec v$, and for all measurement settings $\vec a, \vec b$ under consideration. In the case of polarisation, as initially considered by Leggett~\cite{leggett}, Eqs.~(\ref{constr_ua}--\ref{constr_vb}) impose that the local observations, conditioned on the ``hidden variables'' $\vec u, \vec v$, should satisfy Malus' law.

Equations~(\ref{decomp_leggett}--\ref{constr_vb}) are the \emph{only} constraints imposed by the assumptions of crypto-nonlocality~\cite{footnote_lambdas}, as defined by Leggett in Ref.~\cite{leggett}. Note that the correlations $P_{\vec u, \vec v}(\alpha, \beta | \vec a, \vec b)$ in the decomposition~\eqref{decomp_leggett} are non-signaling, and that no time-ordering of Alice and Bob's measurements needs to be specified~\cite{leggett_NatPhys}. Only Alice and Bob's local marginals $M^A_{\vec u, \vec v}(\vec a, \vec b)$ and $M^B_{\vec u, \vec v}(\vec a, \vec b)$ are constrained by the assumptions of crypto-nonlocality, through Eqs.~(\ref{constr_ua}--\ref{constr_vb}). Nothing is said about the correlation terms $C_{\vec u, \vec v}(\vec a, \vec b) := \sum_{\alpha,\beta} \alpha \beta \, P_{\vec u, \vec v}(\alpha, \beta | \vec a, \vec b)$, which can in particular still make the correlations $P_{\vec u, \vec v}(\alpha, \beta | \vec a, \vec b)$---and hence $P(\alpha, \beta | \vec a, \vec b)$---violate Bell's assumption of local causality~\cite{bell_book} (see below).

It is worth emphasizing also that the crypto-nonlocality constraints~(\ref{decomp_leggett}--\ref{constr_vb}) are defined for specific measurement settings $\vec a, \vec b \in {\cal S}^2$. This is in stark contrast with Bell's local causality assumption for instance, where the measurement settings are just arbitrary labels; this is however analogous to the case where one asks whether a given 2-qubit correlation is compatible with a separable state, as the measurement settings must in general also be specified. 
Note also that the correlations obtained from a quantum state can be compatible with the crypto-nonlocality constraints for certain measurement settings, but may cease to satisfy them if more settings are considered.

As Leggett indeed showed, quantum theory predicts certain correlations which do not satisfy the constraints~(\ref{decomp_leggett}--\ref{constr_vb}); the canonical example is the correlation obtained from the singlet state $\ket{\Psi^-} = \frac{1}{\sqrt{2}}(\ket{01}{-}\ket{10})$, for which $P(\alpha,\beta | \vec a, \vec b) = \frac{1}{4}(1{-}\alpha \beta \, \vec a \!\cdot\! \vec b)$.
One can in fact derive, from the constraints~(\ref{decomp_leggett}--\ref{constr_vb}) and the non-negativity of probability distributions \emph{only}, and for some specific measurement settings, so-called \emph{Leggett inequalities} which can be violated by quantum theory and can be tested experimentally~\cite{leggett,leggett_vienna,leggett_vienna2,leggett_PRL,leggett_NatPhys,leggett_Rai_et_al}. All experiments to date~\cite{leggett_vienna,leggett_vienna2,leggett_PRL,leggett_NatPhys,leggett_OAMs,leggett_neutrons} have shown (up to a few loopholes) a violation of Leggett's crypto-nonlocality, and have been in agreement with quantum predictions.

\medskip

\paragraph{Leggett's crypto-nonlocality vs Bell's local causality.---}

It is quite natural to compare the constraints imposed by the assumptions of crypto-nonlocality to those of Bell's local causality assumption~\cite{bell_book}---a very natural assumption to explain correlations between distant events, but famously incompatible with quantum correlations.

As it turns out, there is in fact no logical relation between the two notions; correlations can independently be compatible or incompatible with Leggett's constraints, and compatible or incompatible with Bell's assumption.
Let me clarify this with the following examples:
\begin{itemize}[leftmargin=6mm]

\item The fully random correlation $P(\alpha,\beta | \vec a, \vec b) = \frac{1}{4}$ for all $\alpha,\beta, \vec a, \vec b$ is compatible both with Leggett's crypto-nonlocality (take e.g. $\vec u$ and $\vec v$ independently and uniformly distributed on ${\cal S}^2$, and define $P_{\vec u, \vec v}(\alpha, \beta | \vec a, \vec b) = \frac{1}{4}(1{+}\alpha \, \vec u \cdot \vec a)(1{+}\beta \, \vec v \cdot \vec b)$) and with Bell's local causality.

\item When all (or sufficiently many and well-chosen) measurement settings $\vec a, \vec b \in {\cal S}^2$ are considered, the singlet state correlations $P(\alpha,\beta | \vec a, \vec b) = \frac{1}{4}(1{-}\alpha \beta \, \vec a \cdot \vec b)$ are incompatible with both Leggett's crypto-nonlocality~\cite{leggett} and Bell's local causality~\cite{bell_book}.

\item However, when the measurements settings under consideration are restricted for instance to the equatorial plane of the Bloch sphere, the singlet state correlations are compatible with Leggett's crypto-nonlocality~\cite{leggett}, but can violate Bell's local causality~\cite{bell_book}.

Another, non-quantum correlation that is compatible with Leggett's assumption but violates Bell's local causality is the ``PR-box'' correlation~\cite{PRbox} $P(\alpha,\beta | \vec a_i, \vec b_j) = \frac{1}{4}(1{+}\alpha \beta \, (-1)^{ij})$, for two measurement settings ($i,j=0,1$) for both Alice and Bob (take $\vec u$ to be orthogonal to both $\vec a_0$ and $\vec a_1$, and $\vec v$ to be orthogonal to both $\vec b_0$ and $\vec b_1$).

\item Lastly, consider deterministic correlations $P(\alpha,\beta | \vec a_i, \vec b) = \delta_{\alpha,1} \, \delta_{\beta,1}$ (where $\delta_{i,j}$ is the Kronecker delta) for at least two different settings $\vec a_0$ and $\vec a_1$ for Alice: such correlations are incompatible with Leggett's crypto-nonlocality (all $P_{\vec u, \vec v}(\alpha, \beta | \vec a, \vec b)$ in the decomposition~\eqref{decomp_leggett} must indeed be such that $M^A_{\vec u, \vec v}(\vec a_0, \vec b) = M^A_{\vec u, \vec v}(\vec a_1, \vec b) = \vec u \cdot \vec a_0 = \vec u \cdot \vec a_1 = 1$, which is impossible for $\vec a_0 \neq \vec a_1$), while they cleary satisfy Bell's local causality assumption. Note also with this example that---in contrast to Bell's local causality---Leggett's crypto-nonlocality can in principle be falsified by considering only one party~\cite{footnote_leggett_one_party}.

\end{itemize}

However artificial these last examples may look (e.g., PR-box correlations are not usually thought of as having Bloch vectors as ``measurement settings''), they illustrate indeed the independence of the two notions of crypto-nonlocality and local causality.
Note that the same observations hold when comparing Leggett's crypto-nonlocality to the concept of quantum steering, a weaker notion of quantum nonlocality~\cite{steering_wiseman} (for the last example, consider e.g. just one setting for Bob, and non-steerability from Alice to Bob).

\medskip
\paragraph{Leggett's crypto-nonlocality vs quantum separability.---}

\hspace{-2\parindent} The next natural question to ask is how Leggett's crypto-nonlocality compares to the notions of quantum separability and quantum entanglement. This was precisely the subject of M. Navascu\'es' note~\cite{leggett_miguel}.

First of all, as noted in Ref~\cite{leggett_miguel}, correlations from 2-qubit separable states obviously satisfy Leggett's axioms~(\ref{decomp_leggett}--\ref{constr_vb}) of crypto-nonlocality.
Reciprocally, one can already see from the previous remarks that some quantum correlations can be compatible with Leggett's axioms, but violate Bell's local causality (cf the third example above). These correlations can therefore not be generated by separable quantum states, which shows that crypto-nonlocality and quantum separability are not equivalent concepts---contrary to the claims in Ref.~\cite{leggett_miguel}.

One could reply that the argument above considers only a limited number of measurement settings on the 2-qubit singlet state; indeed, as noted before, when all measurements are allowed the statistics of the singlet state are incompatible with Leggett's assumption of crypto-nonlocality. Could it then be that when all measurement settings are allowed, the correlations from any entangled state are incompatible with Leggett's constraints?

The answer is negative. I show in the Appendix how to construct an explicit ``Leggett model'' that reproduces the correlations $P(\alpha,\beta | \vec a, \vec b) = \frac{1}{4}(1{-}\alpha \beta \, V \, \vec a \cdot \vec b)$ of 2-qubit Werner states~\cite{werner} $\varrho_V = V \, \ket{\Psi^-}\!\bra{\Psi^-} + (1-V) \, \frac{\one}{4}$ for all $V \in [0,\frac{1+1/\sqrt{2}}{2}]$, and for all settings $\vec a, \vec b \in {\cal S}^2$. Now, Werner states are entangled for $V > 1/3$: for $1/3 < V \leq \frac{1+1/\sqrt{2}}{2} \simeq 0.85$, they thus provide an example of entangled states that are compatible with Leggett's crypto-nonlocality for all projective measurements~\cite{footnote_werner,footnote_POVM}. This reinforces the claim that crypto-nonlocality and quantum separability are genuinely different notions: quantum separability implies crypto-nonlocality, but not reciprocally---and one cannot ``[regard] any two-qubit entanglement witness [...] as a Leggett inequality''~\cite{leggett_miguel}.

\medskip
\paragraph{On Navascu\'es' argument~\cite{leggett_miguel}.---}

As mentioned, my claims above appear to contradict those of Ref.~\cite{leggett_miguel}. To understand this discordance, note that Ref.~\cite{leggett_miguel} deals more with a possible \emph{physical interpretation} of Leggett's notion of crypto-nonlocality than with that notion itself, rigorously defined through Eqs.~(\ref{decomp_leggett}--\ref{constr_vb}).
Invoking indeed the physical axioms proposed in Ref.~\cite{leggett_vienna}, Navascu\'es actually introduces additional assumptions, which he claims are ``reasonable''~\cite{leggett_miguel}, but which are only inspired by a particular interpretation of Leggett's notion---e.g., that the subensembles of photons corresponding to the ``hidden variables'' $\vec u, \vec v$ \emph{have} definite polarisations described by $\vec u$ and $\vec v$ (rather than ``\emph{locally behave as if they had} definite polarisations'', which is all the constraints~(\ref{constr_ua}--\ref{constr_vb}) suggest and still allows for non-trivial correlation coefficients); or that Bob's physical state, for a given measurement result of Alice, is also a mixture of photons that \emph{have} a definite polarisation. These additional constraints do not however strictly speaking follow from Leggett's axioms~(\ref{decomp_leggett}--\ref{constr_vb}).
By further imposing such constraints on crypto-nonlocality, Navascu\'es actually distorts Leggett's notion to the point where it is merely reduced, as he shows, to quantum separability. Without these additional constraints, such a conclusion could however not be reached, as I highlighted above. It is precisely because Leggett freed himself from certain ``physical intuitions'' and did not impose the additional constraints suggested by Navascu\'es, that the notion of crypto-nonlocality he introduced is not just ``a bizarre reformulation of quantum separability''~\cite{leggett_miguel}.

Nevertheless, the merit of Ref.~\cite{leggett_miguel} is that it questions the physical motivation of Leggett's crypto-nonlocality, and shows that if certain ``physically reasonable'' assumptions are added, it becomes somehow trivial. Hence, in order to remain non-trivial, one may argue that Leggett's notion must incorporate quite unnatural features. Whether an explanation of correlations based on crypto-nonlocality would then be satisfying is indeed debatable---but this question is beyond the scope of the present note.

\medskip

\paragraph{Conclusion.---}

The objective of this note was to clarify the fact that Leggett's notion of crypto-nonlocality is quite different from those of local causality and quantum separability. It is only when crypto-nonlocality is further constrained by additional assumptions, that it can, as shown in~\cite{leggett_miguel}, become equivalent to quantum separability.

The concept of crypto-nonlocality is certainly not as important for our understanding of quantum correlations as that of local causality for instance (as its physical motivation may not be as clear), and its falsifiability by quantum theory does not so deeply and fundamentally challenge our conception of the physical world than the violation of Bell inequalities. Still, the study of Leggett's assumptions has already inspired some interesting results, e.g. on the predictive power and the completeness of quantum theory~\cite{leggett_NatPhys,colbeck_renner}, and may generate more insights in the future. I see no reason to ``hope to put an end to this line of research in the Foundations of Quantum Physics''~\cite{leggett_miguel}.

\medskip
\paragraph{Acknowledgments.---}

This work was supported by a University of Queensland postdoctoral research fellowship.

\appendix

\subsubsection{\textbf{Appendix}}

\paragraph{Bipartite, binary-outcome correlations.---}

A convenient way to write any bipartite correlation $P_{\vec u, \vec v}(\alpha, \beta | \vec a, \vec b)$ (as in the decomposition~\eqref{decomp_leggett}) with binary outcomes $\alpha, \beta = \pm 1$ is
\ba
P_{\vec u, \vec v}(\alpha, \beta | \vec a, \vec b) &=& \frac{1}{4} \Big(1 + \alpha \, M^A_{\vec u, \vec v}(\vec a, \vec b) \nonumber \\[-1mm]
&& \hspace{6mm} + \beta \, M^B_{\vec u, \vec v}(\vec a, \vec b) + \alpha \beta \, C_{\vec u, \vec v}(\vec a, \vec b) \Big), \qquad \label{useful_decomp}
\ea
where $M^A_{\vec u, \vec v}(\vec a, \vec b) = \sum_{\alpha,\beta} \alpha \, P_{\vec u, \vec v}(\alpha, \beta | \vec a, \vec b)$ and $M^B_{\vec u, \vec v}(\vec a, \vec b) = \sum_{\alpha,\beta} \beta \, P_{\vec u, \vec v}(\alpha, \beta | \vec a, \vec b)$ are the marginals on Alice and Bob's side, while $C_{\vec u, \vec v}(\vec a, \vec b) = \sum_{\alpha,\beta} \alpha \beta \, P_{\vec u, \vec v}(\alpha, \beta | \vec a, \vec b)$ is the correlation coefficient.
The constraint that the probabilities $P_{\vec u, \vec v}(\alpha, \beta | \vec a, \vec b)$ must be non-negative, for all $\alpha$ and $\beta$, is equivalent to
\ba
&& \hspace{-1cm} -1 + | M^A_{\vec u, \vec v}(\vec a, \vec b) + M^B_{\vec u, \vec v}(\vec a, \vec b) | \nonumber \\
&& \leq \ C_{\vec u, \vec v}(\vec a, \vec b) \ \leq \ 1 - | M^A_{\vec u, \vec v}(\vec a, \vec b) - M^B_{\vec u, \vec v}(\vec a, \vec b) |. \label{constr_pos}
\ea

\medskip
\paragraph{An explicit Leggett model for 2-qubit Werner states of visibility $V \leq \frac{1+1/\sqrt{2}}{2}$.---}

Let us choose $\vec v = -\vec u$, with $\vec u$ uniformly distributed on the Bloch sphere ${\cal S}^2$, and assume that the correlations $P_{\vec u, \vec v}(\alpha, \beta | \vec a, \vec b)$ satisfy the crypto-nonlocality constraints~(\ref{constr_ua}--\ref{constr_vb}). After integrating over $\vec u$, we find, for all $\vec a, \vec b$,
\ba
M^A(\vec a, \vec b) &\! := \! & \int_{{\cal S}^2} \! M^A_{\vec u, -\vec u}(\vec a, \vec b) \, \frac{d \vec u}{4\pi} \, = \, \int_{{\cal S}^2} \! (\vec u \cdot \vec a) \, \frac{d \vec u}{4\pi} \, = \, 0, \nonumber \\
M^B(\vec a, \vec b) &\! := \! & \int_{{\cal S}^2} \! M^B_{\vec u, -\vec u}(\vec a, \vec b) \, \frac{d \vec u}{4\pi} \, = \,\int_{{\cal S}^2} \! \!(- \vec u \cdot \vec b) \, \frac{d \vec u}{4\pi} \, = \, 0, \nonumber
\ea
which are the marginals expected for the correlations of the Werner state $\varrho_V = V \, \ket{\Psi^-}\!\bra{\Psi^-} + (1-V) \, \frac{\one}{4}$. In order to obtain the full correlations of the Werner state, we also need the correlation coefficient to be
\ba
C(\vec a, \vec b) &\! := \! & \int_{{\cal S}^2} C_{\vec u, -\vec u}(\vec a, \vec b) \, \frac{d \vec u}{4\pi} \, = \, - V \, \vec a \cdot \vec b. \label{constr_C_Vab}
\ea

Now, Eq.~\eqref{constr_pos} writes here
\ba
&& \hspace{-1cm} -1 + | \vec u \cdot (\vec a - \vec b) | \ \leq \ C_{\vec u,-\vec u}(\vec a, \vec b) \ \leq \ 1 - | \vec u \cdot (\vec a + \vec b) |. \label{constr_C}
\ea
After integrating it over the values of $\vec u$ (with $\int_{{\cal S}^2} |\vec u \cdot (\vec a \pm \vec b)| \, \frac{d \vec u}{4\pi} = \frac{1}{2}|| \vec a \pm \vec b || = \sqrt{\frac{1 \pm \vec a \cdot \vec b}{2}}$) and using the constraint~\eqref{constr_C_Vab}, this gives the necessary condition that
\ba
&-1 + \sqrt{\frac{1 - \vec a \cdot \vec b}{2}} \, \leq \, -V \, \vec a \cdot \vec b \, \leq \, 1 - \sqrt{\frac{1 + \vec a \cdot \vec b}{2}} \, , \quad
\ea
which indeed holds for all values of $\vec a \cdot \vec b \ \in [-1,1]$ (i.e. all unit vectors $\vec a, \vec b \in {\cal S}^2$) when $0 \leq V \leq \frac{1+1/\sqrt{2}}{2}$.

For such values of $V$, it is always possible to find functions $C_{\vec u, -\vec u}(\vec a, \vec b)$ satisfying~\eqref{constr_C_Vab} and~\eqref{constr_C} for all $\vec a, \vec b$ and $\vec u$; one can choose for instance
\ba
C_{\vec u, -\vec u}(\vec a, \vec b) &=& p_{\vec a, \vec b}^- \, \big[-1 + | \vec u \cdot (\vec a - \vec b) | \, \big] \nonumber \\
&& \ + \ p_{\vec a, \vec b}^+ \, \big[ \, 1 - | \vec u \cdot (\vec a + \vec b) | \, \big] \nonumber \\[1mm]
\text{with} \ \ p_{\vec a, \vec b}^\pm &=& \frac{1 - \sqrt{\frac{1 \mp \vec a \cdot \vec b}{2}} \mp V \, \vec a \cdot \vec b}{2 - \sqrt{\frac{1 + \vec a \cdot \vec b}{2}} - \sqrt{\frac{1 - \vec a \cdot \vec b}{2}}} \, , \ \
\ea
such that for all $\vec a, \vec b$, $p_{\vec a, \vec b}^+ + p_{\vec a, \vec b}^- = 1$ and (for $V \leq \frac{1+1/\sqrt{2}}{2}$) $p_{\vec a, \vec b}^\pm \geq 0$. This, together with the constraints~(\ref{constr_ua}--\ref{constr_vb}) on the marginals, then defines valid (i.e. non-negative) probabilities $P_{\vec u, \vec v}(\alpha, \beta | \vec a, \vec b)$ through Eq.~\eqref{useful_decomp}, and thus provides an explicit Leggett model for all projective measurements on 2-qubit Werner states $\varrho_V$ of visibility $V \leq \frac{1+1/\sqrt{2}}{2}$.


\begin{thebibliography}{99}
\bibitem{footnote_leggett_one_party} Note however that the deterministic correlations $P(\alpha,\beta | \vec a_i, \vec b) = \delta_{\alpha,1} \, \delta_{\beta,1}$ considered in the example cannot be obtained quantum mechanically from qubit measurements with $\vec a_0 \neq \vec a_1$; hence, it cannot be used to falsify Leggett's crypto-nonlocality assumption in a (quantum) experiment where the implementation of the measurements is trusted.
\bibitem{footnote_POVM} One may also ask whether the correlations obtained from Positive Operator-Valued Measurements (POVMs) on Werner states satisfy a generalized version of crypto-nonlocality, that would precisely also consider general POVMs---the constraints~(\ref{decomp_leggett}--\ref{constr_vb}) can indeed straightforwardly be generalized to POVMs, although I am not aware of any work in this direction. I conjecture that there is indeed a range of entangled Werner states that would still be compatible with that generalized version of crypto-nonlocality (analogously, POVMs on Werner states admit a locally causal model for $V \leq \frac{5}{12}$~\cite{barrett}).
\end{thebibliography}
\end{document}